\documentclass[9pt,twocolumn,twoside,lineno]{pnas-new}

\templatetype{pnasresearcharticle} 

\usepackage{pdfpages}

\newcommand\icarus{Icarus}
\newcommand\mnras{MNRAS}
\newcommand\aj{{AJ}}
\newcommand\araa{{Ann.~Rev.~Astr.~Astrophys.}}

\title{Pluto near the edge of chaos}

\author[a,1]{Renu Malhotra}
\author[b,c]{Takashi Ito}

\affil[a]{Lunar and Planetary Laboratory, The University of Arizona, 1629 E University Blvd, Tucson, AZ 85721}
\affil[b]{Center for Computational Astrophysics, National Astronomical Observatory of Japan, Osawa 2--21--1, Mitaka, Tokyo 181--8588, Japan}
\affil[c]{Planetary Exploration Research Center, Chiba Institute of Technology, 2--17--1 Tsudanuma, Narashino, 275--0016, Chiba, Japan}

\leadauthor{Malhotra} 

\significancestatement{
The dwarf planet, Pluto, has stirred the imagination of the public and of planetary scientists due to its many unusual properties. Amongst these properties is its Neptune-crossing orbit whose stability is owed to an orbital resonance with Neptune. Less well understood is the role of the other planets. We demonstrate that the orbital architecture of the giant planets lies within a narrow niche in which Pluto-like orbits are practically stable on gigayear timescales, whereas nearby are strongly chaotic orbits.
Pluto is witness to the dynamical history of the solar system; quantifying its proximity to strong chaos and  dynamical instability  can enable quantitative constraints on its own dynamical history as well as that of the solar system.
}

\authorcontributions{
RM designed the research, carried out parts of it (a portion of the numerical simulations and the entire analytic and conceptual formulation), and wrote the paper. TI carried out parts of the research (a major portion of the numerical simulations) and contributed to writing the paper.}
\authordeclaration{
The authors declare no competing interest.}
\correspondingauthor{\textsuperscript{1}To whom correspondence should be addressed. E-mail: malhotra@arizona.edu}


\begin{abstract}
Many of the unusual properties of Pluto's orbit are widely accepted as evidence for the orbital migration of the giant planets in early solar system history. However, some properties remain an enigma. Pluto's long term orbital stability is supported by two special properties of its orbit that limit the location of its perihelion in azimuth and in latitude. We revisit Pluto's orbital dynamics with a view to elucidating the individual and collective gravitational effects of the giant planets on constraining its perihelion location. While the resonant perturbations from Neptune account for the azimuthal constraint on Pluto's perihelion location, we demonstrate that the long term and steady persistence of the latitudinal constraint is possible only in a narrow range of additional secular forcing which arises fortuitously from the particular orbital architecture of the other giant planets. Our investigations also find that Jupiter has a largely stabilizing influence whereas Uranus has a largely destabilizing influence on Pluto's orbit. Overall, Pluto's orbit is rather surprisingly close to a zone of strong chaos. 
\end{abstract}

\dates{This manuscript was compiled on February 18, 2022}
\doi{\url{www.pnas.org/cgi/doi/10.1073/pnas.XXXXXXXXXX}}

\begin{document}
\renewcommand{\today}{February 18, 2022} 

\maketitle

\dropcap{P}luto's orbit is significantly eccentric and its orbit plane is inclined 17 degrees to the solar system's invariable plane.
Its eccentric orbit overlaps that of Neptune's, so much so that for approximately two decades of its 248 year long orbital period it is closer to the Sun than is Neptune; Pluto's most recent perihelion passage {closer to the Sun than Neptune} was observed during the period 1979 to 1999.
In general,
most such planet-crossing orbits have very short dynamical stability time because planetary close encounters cause large, destabilizing perturbations. 
Numerical propagation of Pluto's orbit shows that Pluto avoids close encounters with Neptune due primarily to two types of librations of its perihelion \citep[see, e.g.,][]{Malhotra:1997}. {[``Libration'' is a term used in celestial mechanics for the oscillation of an angular variable or a combination of angular variables.]}
The first and most consequential for Pluto's dynamical stability is the libration of its perihelion longitude about a center $\pm 90^\circ$ away from Neptune's ecliptic longitude. {[Here ``ecliptic'' refers to the standard plane of reference for orbits in the solar system.]} This libration, which has a period of about 20,000 years, is associated with Pluto's 3/2 mean motion resonance with Neptune (Pluto's orbital period is 1.5 times Neptune's orbital period), and is characterized by the libration of the critical resonant angle,
\begin{equation}
    \phi=3\lambda-2\lambda'-\varpi,
    \label{e:phi}
\end{equation}
where $\lambda$ and $\varpi$ denote Pluto's mean longitude and its longitude of perihelion, and $\lambda'$ denotes Neptune's mean longitude.
The libration of $\phi$ about a center at 180$^\circ$, with an amplitude of $80^\circ$--$86^\circ$, ensures that at the times when Pluto crosses Neptune's orbit its spatial location is far removed from Neptune's, more than 45$^\circ$ in ecliptic longitude. We call this an azimuthal libration.

The second is the libration of its perihelion in the third dimension: at perihelion Pluto's location oscillates about a high 
latitude, well above the plane of the {other} planets. We call this  a latitudinal libration.
In terms of orbital elements, it is characterized by the libration of Pluto's argument of perihelion, $\omega$, about a center at $90^\circ$, an amplitude of $24^\circ$--$27^\circ$ and a period of $\sim$~4 Myr.
[The argument of perihelion is the angular distance of Pluto's perihelion from its longitude of ascending node.
The latter is usually reported in the ecliptic plane although the dynamically relevant reference plane is closer to Neptune's orbit plane or to the Solar system's invariable plane.]
The libration of $\omega$ has the effect of elevating Pluto's minimum distance of approach to Neptune and to the other giant planets, thereby increasing its orbital stability.  {A visualization of Pluto's spatial perihelion librations can be found in \cite{Zaveri:2021}.}

After several foundational studies on Pluto's dynamics in the 1960s and 1970s \citep[e.g.,][]{Cohen:1965,Williams:1971,Nacozy:1978b}, very long numerical orbit propagations of sufficient accuracy became possible with advanced digital computers in the late 1980s. {These} shed more light on Pluto's engagement with the giant planets in multiple resonances and its potential for chaotic orbital evolution on very long timescales \citep{Sussman:1988,Milani:1989}. Pluto is one of the first examples in solar system dynamics whose chaotic nature was unveiled. Sussman \& Wisdom \citep{Sussman:1988} propagated the orbital motion of the outer four giant planets and Pluto for 845 million years, and found that its nearby trajectories diverge exponentially with an $e$-folding time of only about 20 million years. Later, Laskar \citep{laskar1990} numerically solved the secular equations of motion of the eight major planets (excluding Pluto), and claimed that the entire solar system is chaotic in the sense that its Lyapunov index is positive. Then, Sussman \& Wisdom \citep{Sussman:1992} carried out a numerical integration including the interactions of all nine planets (including Pluto, which was considered a planet at that time), and found that the planetary system as a whole is chaotic and its Lyapunov $e$-folding time is only about 4 million years. However, additional long term numerical solutions for the solar system planets established that, within the highest fidelity solar system model, the orbits of the outer planets, including Pluto's, are practically stable on multi-gigayear timescales, both in the past and in the future~\citep{Kinoshita:1996,Ito:2002,Laskar:2008}.  The detection of positive Lyapunov exponents notwithstanding, Pluto's and the planets' perihelion and aphelion distances and their latitudinal variations remain well bounded on multi-gigayear timescales, {indicating that the chaos detected in the above investigations is very weak indeed.}

The period of 1990s--2000s also saw the advancement of the hypothesis of resonance sweeping and capture of Pluto during an early epoch of giant planet migration \citep{Malhotra:1993}. This hypothesis provides a plausible account of Pluto's eccentric resonant orbit within the physical and dynamical processes of the early evolution of the solar system.  {In turn,} Pluto's orbital properties provide quantitative constraints on 
the magnitude and speed of the early migration of the giant planets and the mass and size of the planetesimal disk left over after planet formation \citep[e.g.][]{Malhotra:1995,Yu:1999,Murray-Clay:2006}. A wide range of solar system data has been identified in support of this hypothesis which has been developed extensively in many recent studies; reviews can be found in \cite{Chiang:2007}, \cite{Nesvorny:2018} and \cite{Malhotra:2019b}.

During the period 2005--2012, deep imaging with the Hubble Space Telescope revealed that Pluto hosts a retinue of four small moons, in addition to its large moon, Charon, discovered previously in 1978 \citep{Weaver:2016}.
In 2015, new data from the \textit{New Horizons} spacecraft's reconnaissance of the Pluto system revealed the surprisingly active geophysical state of Pluto \citep{Stern:2018}. These new discoveries have added to the list of puzzles presented by this distant dwarf planet. 

While its origin story is now understood in broad terms, Pluto and its orbital dynamics in the current solar system still present many unsolved problems. These puzzles prompt us to seek to better understand its orbital stability. The present work is a step towards this goal.
We revisit the topic of Pluto's orbital dynamics, with a view to understand better the collective and individual influence of the giant planets on the stability of Pluto's orbit, as manifested in the librations of its critical resonant angle, $\phi$, and of its argument of perihelion, $\omega$.  We use tailored numerical experiments to identify new details of both the secular and non-secular effects of the giant planets' perturbations on Pluto to learn more about the dynamical neighborhood in which Pluto orbits.

\section*{Numerical experiments}

\subsection*{N-body simulations}\label{ss:nbody}

We carried out numerical simulations of Pluto's orbital evolution for up to 5 Gyr with eight different combinations of the perturbing giant planets: 
Neptune only {(hereafter referred to as} \texttt{-{}-{}-NP}),
Uranus$+$Neptune (\texttt{-{}-UNP}), 
Saturn$+$Neptune (\texttt{-S-NP}), 
Jupiter$+$Neptune (\texttt{J-{}-NP}), 
Saturn$+$Uranus$+$Neptune (\texttt{-SUNP}), 
Jupiter$+$Uranus$+$Neptune (\texttt{J-UNP}), 
Jupiter$+$Saturn$+$Neptune (\texttt{JS-NP}), 
and the highest fidelity model with Jupiter$+$Saturn$+$Uranus$+$Neptune (\texttt{JSUNP}).
It must be emphasized that we do not expect these experiments to accurately identify the direct effects of each individual giant planet on Pluto, because the perturbations of each giant planet are not simply a sum of perturbations of each individual perturber.
The giant planets' orbital evolution itself depends upon their mutual interactions. For example, the evolution of Neptune in the Jupiter$+$Neptune model (\texttt{J-{}-NP}) is slightly different than in the Uranus$+$Neptune model (\texttt{-{}-UNP}). Nevertheless, we will see that we gain useful insights with these experiments.  

\begin{figure*}[!t]\centering
\includegraphics[scale=0.9]{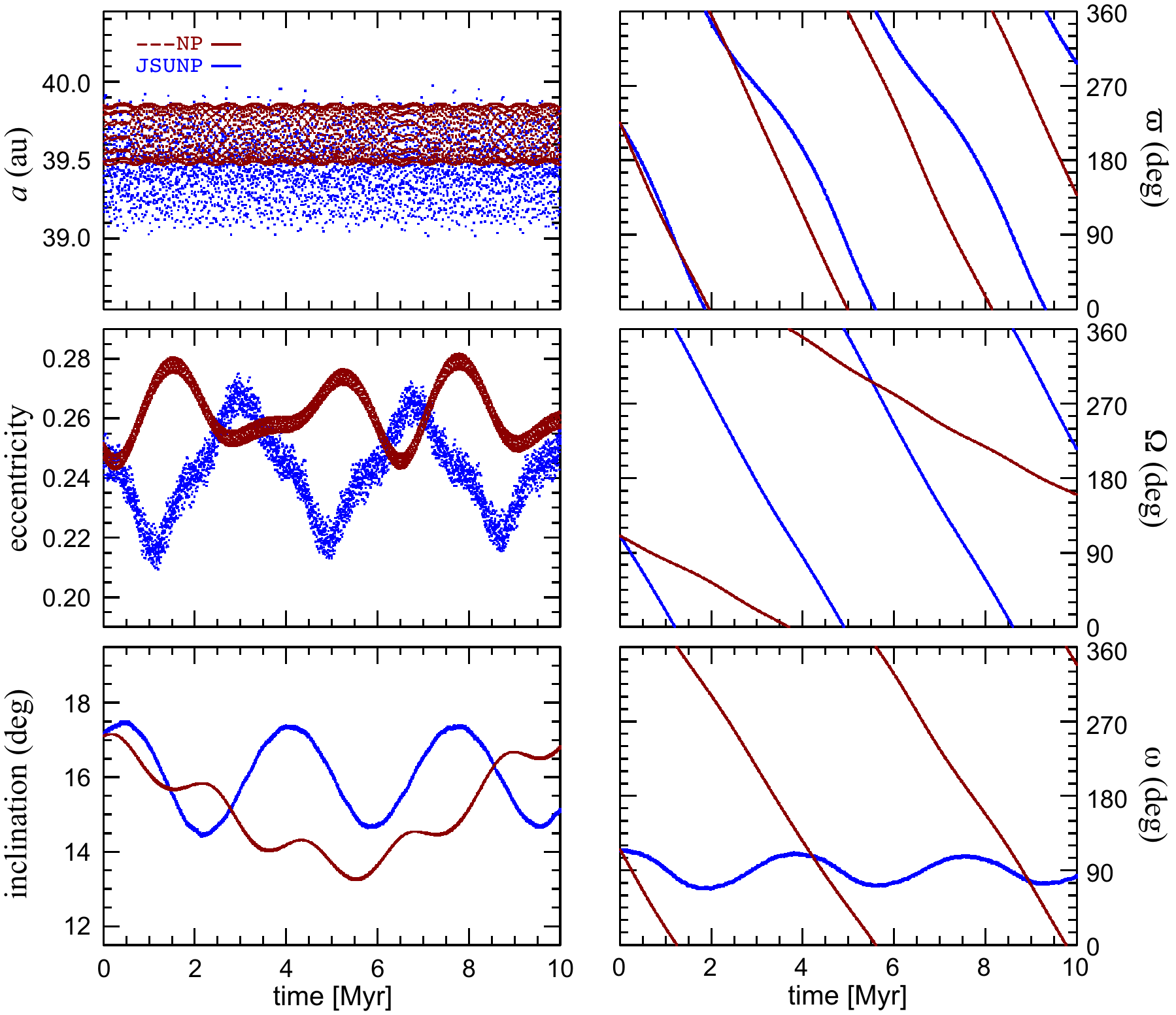}
\caption{Pluto's orbital elements for 10 Myr computed with two different models: the three body model of Sun, Neptune and Pluto ({\texttt{-{}-{}-NP}}; shown in dark red color) and the highest fidelity model of Sun, Jupiter, Saturn, Uranus, Neptune and Pluto ({\texttt{JSUNP}}; shown in blue). Pluto is treated as a massless object in all these models.  {From top to bottom, the  panels on the left plot the semimajor axis, eccentricity and inclination; the panels on the right plot the longitude of perihelion, $\varpi$, the longitude of ascending node, $\Omega$, and the argument of perihelion, $\omega$.
The output interval for the plots is 2.5 Kyr.}
}
\label{f:fig1}
\end{figure*}

We obtained the heliocentric orbital elements of the major planets and Pluto from \href{https://ssd.jpl.nasa.gov/horizons/app.html#/}{JPL Horizons System}, and they are the values as of 2021 March 19 00:00:00 TDB (Barycentric Dynamical Time).
The position and the velocity of each planet are those of its barycenter (including its satellite system).
Pluto's position and velocity are also those of the barycenter of the Pluto system.
We obtained the masses of the planets from JPL's DE245 \citep[e.g.][]{fukushima1995}.
In the simulations, we regard Pluto as a massless particle.
For the numerical orbit propagation, we employed the second-order regularized mixed variable symplectic integrator
based on the Wisdom--Holman symplectic mapping \citep{Wisdom:1991}
implemented as a part of the \textsc{swift} package \citep{levison1994}. 
[The code and data availability are described in the Supplementary Information.]
We adopted a basic stepsize of ten days.
 {For computational efficiency, we set an outer cutoff distance of 100 au so as to cease numerical propagation of Pluto-like particles that are ejected from Neptune's 3/2 mean motion resonance.}
For checking the accuracy, we also carried out some of the same orbit propagations with
the fourth-order standard symplectic integrator that splits the Hamiltonian just into kinetic energy and potential energy terms \citep[e.g.][]{yoshida1990}.
We confirmed that both the methods yield similar output in terms of Pluto's dynamical characteristics that we discuss in this work.
For illustration, we plot in Figure~1 
the time evolution of Pluto's orbital elements for two models, the simplest model with Neptune as the sole perturber (\texttt{-{}-{}-NP}) and the highest fidelity model with all four giant planet perturbers (\texttt{JSUNP}); the contrast in some properties of Pluto's orbital evolution between these two models is quite stark, and we discuss these differences later on.

\begin{figure*}[!t]\centering
\includegraphics[scale=0.9]{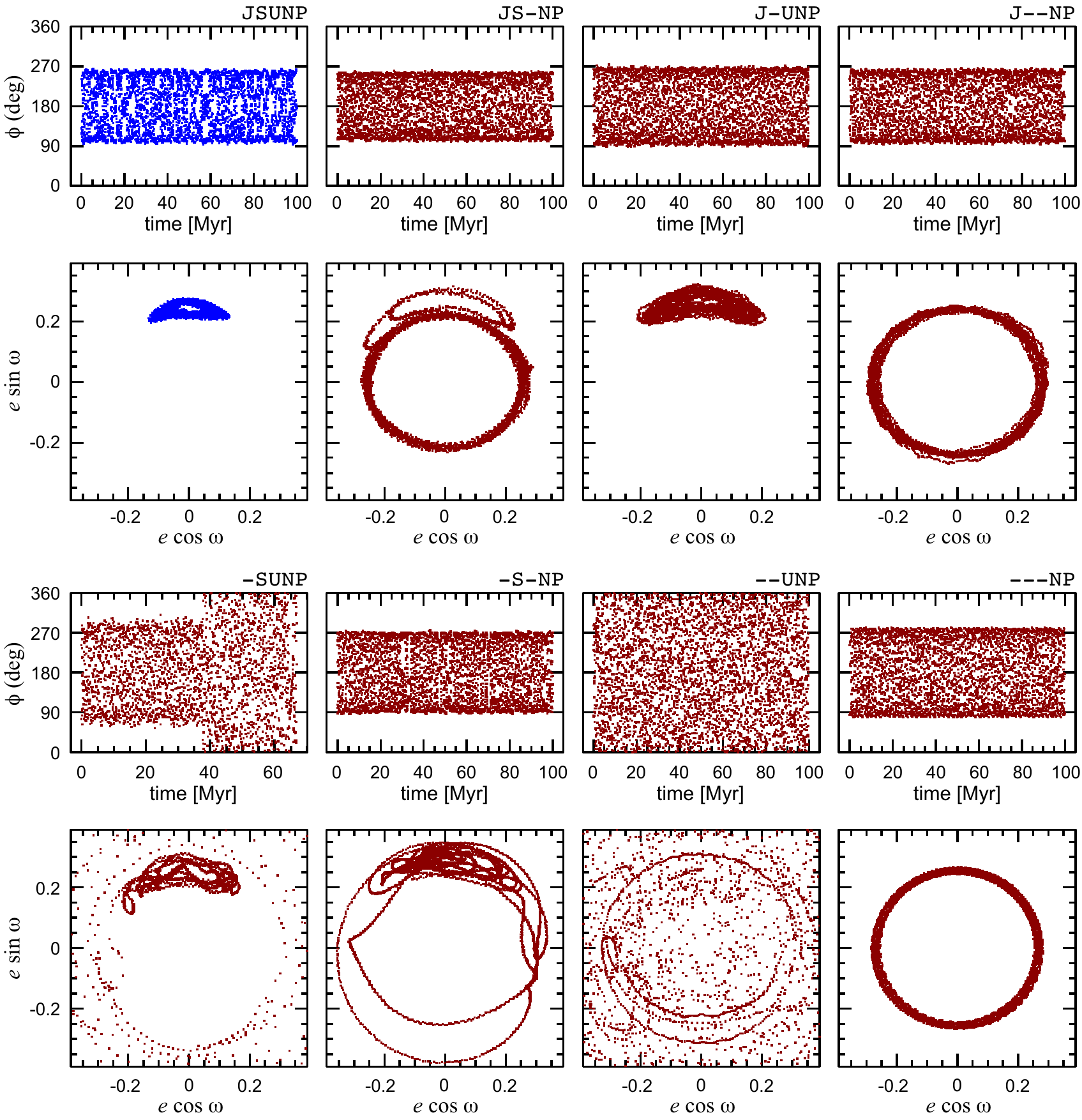}
\caption{The evolution over 100 Myr of Pluto's critical resonant angle $\phi$ (\eqref{e:phi}), and argument of perihelion $\omega$, in numerical simulations with different sets of perturbers. The top left (in blue) is for the highest fidelity model including all four giant planets and Pluto (\texttt{JSUNP}), the other cases are shown in dark red, with the planets in the model indicated in the legend above each pair of panels (for example, \texttt{-{}-UNP} indicates the model with Uranus, Neptune and Pluto). Note that in the case of model \texttt{-SUNP}, the simulation lasts only about 67 Myr, ending because Pluto travelled too far from the Sun (heliocentric distance $r > 100$ au) on a very extended orbit  {no longer confined to Neptune's 3/2 mean motion resonance.
The output interval in the plots is 25 Kyr.}
}
\label{f:nbodsA}
\end{figure*}%

\begin{figure*}[!t]\centering
\includegraphics[scale=0.9]{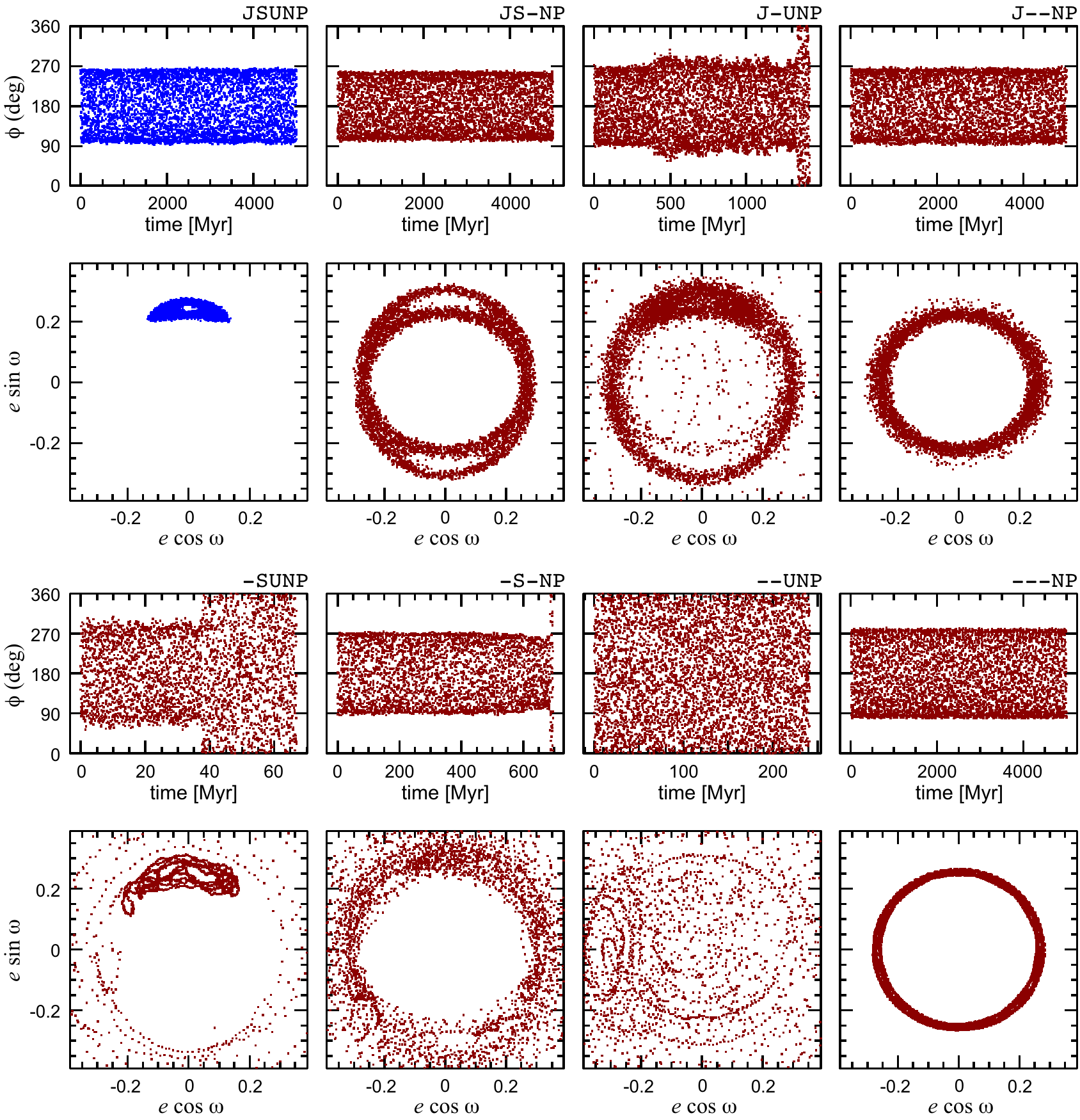}
\caption{Similar to Figure~\ref{f:nbodsA}, but for the evolution up to 5 Gyr. The output interval in the plots is 1 Myr for those models in which Pluto's stability time is at least 5 Gyr; for the cases with shorter stability times, the output interval is shorter, in the range 20 Kyr to 250 Kyr.
}
\label{f:nbodsB}
\end{figure*}

For each of the eight models, we examined the behavior of 
 {$\phi$ and $\omega$} by making two plots: a time series plot of $\phi(t)$ and a polar plot of {$(e \cos \omega, e \sin \omega)$.}
These pairs of plots are shown in Figure~\ref{f:nbodsA} and Figure~\ref{f:nbodsB} for each of the eight models. In Figure~\ref{f:nbodsA} we plot only the first 100 Myr of evolution, and in Figure~\ref{f:nbodsB} we plot the evolution for up to 5 Gyr in the eight models.  By examining these results closely, we observe the following. 
\begin{enumerate}
\item The simplest model with Neptune as the sole perturber (model \texttt{-{}-{}-NP}) maintains the libration of $\phi$ but not of $\omega$; the latter  {does not librate, but} undergoes fairly smooth rotations in this model; these  {rotations} persist on gigayear timescales in this model.
\item  Adding Uranus to the simplest model (\texttt{-{}-UNP}) destroys the libration of $\phi$ within a few megayears, and it also causes $\omega$ to evolve  {irregularly, with intermittent librations and  rotations}; the dynamical lifetime of Pluto is only a few hundred megayears in this case. 
However, adding either Jupiter or Saturn to the simplest model (\texttt{J-{}-NP} and \texttt{-S-NP}, respectively) yields longer dynamical lifetimes by stabilizing the librations of $\phi$ (but not of $\omega$). With the addition of Saturn, the dynamical lifetime gain is modest, only a factor of 2--3, but Jupiter's addition increases the dynamical lifetime to at least 5 Gyr. We find that in the latter case (\texttt{J-{}-NP}), $\omega$ undergoes fairly smooth rotations, but it is significantly slowed compared to the simplest case (\texttt{-{}-{}-NP}). 
\item Of the three models with three perturbing planets, the model without Jupiter (\texttt{-SUNP}) yields chaotic evolution of both $\phi$ and $\omega$ and a dynamical lifetime less than 100 Myr.  {But} the models with Jupiter (\texttt{J-UNP} and \texttt{JS-NP}) have dynamical lifetimes exceeding a gigayear. These latter two models support librations of $\phi$ and $\omega$ for at least ten million years. We found that in the \texttt{JS-NP} model $\omega$ slips from libration to chaotic evolution (intermittent libration and rotation) in less than 100 Myr while $\phi$ remains in steady libration for 5 Gyr. In the \texttt{J-UNP} model, both $\omega$ and $\phi$ slip into chaotic evolution on timescales of a few hundred megayears, and Pluto is ejected from the 3/2 mean motion resonance in less than 2 Gyr.
\item All models yield libration of $\phi$ on at least $\sim1$~Myr, albeit with differences in libration amplitudes. Two models, \texttt{-{}-UNP} and \texttt{-SUNP}, do not support steady librations of $\phi$ on longer than 1 Myr timescales. These two models can be reasonably considered the two most unstable cases. With different integration schemes, such as the Bulirsch--Stoer extrapolation method \citep{hairer1993}, the trajectories of Pluto in these two models also diverge visibly over just $\sim$~1~Myr.  {The divergence of solutions with different integrators is another symptom of the strongly chaotic behavior in those models.}
\item Notably, only the highest fidelity model (\texttt{JSUNP}) yields the steady libration of both $\phi$ and $\omega$ on gigayear timescales.
\end{enumerate}

\subsection*{Modified restricted three body model}\label{s:secular}

As previous studies have shown, and the numerical experiments in the previous section have confirmed, Neptune's resonant perturbations 
maintain the libration of the resonant angle, $\phi$.
But the additional libration of $\omega$ cannot occur with Neptune's perturbations alone.
Previous semi-analytic studies concluded that the secular effects of the three inner giant planets, Jupiter--{Saturn--}Uranus, are critical to support the libration of Pluto's $\omega$ \citep{Nacozy:1978b}. We briefly outline this mechanism, and then follow up with numerical experiments with a modified three body model to quantitatively examine this hypothesis.

The condition for the libration of Pluto's $\omega$ is that its time-averaged rate must vanish, i.e., $\left< \dot{\omega} \right> = 0$.
Considering that $\omega=\varpi-\Omega$, this requires that Pluto's average apsidal rate, $\dot\varpi$, matches its average nodal rate, $\dot\Omega$. 
With Neptune's perturbations alone, we find (from the numerical solution of the  {Sun--Neptune--Pluto} three body model, i.e., the \texttt{-{}-{}-NP} model)
that Pluto's apsidal rate is
  $\dot\varpi_\mathrm{(only\; Neptune)} \simeq -11.3 \times 10^{-5}$ ~deg yr$^{-1}$,
and its nodal rate is
  $\dot\Omega_\mathrm{(only\; Neptune)} \simeq -3    \times 10^{-5}$ ~deg yr$^{-1}$ (see Figure~1). 
The apsidal rate forced by Neptune can be understood as being of two parts, one part from the secular perturbations (which contributes a precession) and another part from the mean motion resonance (which contributes a regression); the latter is dominant and leads to an overall net regression for Pluto in the \texttt{-{}-{}-NP} model. The secular forcing by the inner three giant planets (Jupiter, Saturn and Uranus) is qualitatively similar in effect to that of an additional quadrupolar potential which contributes a positive apsidal rate and a negative nodal rate.
As we will see quantitatively with the numerical analyses below, the secular effects from the inner three giant planets are just enough that Pluto's nodal and apsidal rates become nearly equal, leading to the near-vanishing of the average rate of Pluto's $\omega$, and thereby supporting its libration.
These effects can be discerned in Figure~1 
in which we can observe that in the model with all four giant planets (\texttt{JSUNP}), Pluto's apsidal regression is smaller while its nodal regression is larger than in the three body model in which Neptune is the sole perturber (\texttt{-{}-{}-NP}).
The apsidal and nodal rates are nearly equal in the \texttt{JSUNP} model, accounting for the near-vanishing of the rate of $\omega$. 

In order to more directly test the above hypothesis, we carry out simulations with a modified restricted three body model in which we included the secular effects of the inner three giant planets as follows. The secular effects of the inner three giant planets can be approximately modelled by replacing each giant planet with a circular ring of radius equal to the planet's semi major axis $a_\mathrm{p}$ and mass equal to the planet's mass $m_\mathrm{p}$. The gravitational potential of a ring at a heliocentric distance $r > a_\mathrm{p}$ and a distance $z$ above the plane of the ring, is given by  \citep[e.g.][]{Jackson:1975book}
\begin{equation}
  V_\mathrm{ring} = -\frac{G m_\mathrm{p}}{r} \left[ 1 + \sum_{k=1}^\infty \left(\frac{a_\mathrm{p}}{r}\right)^{2k} P_{2k}(0)P_{2k} \left( \frac{z}{r}\right) \right], 
\label{e:Vring}
\end{equation}
where $G$ is the universal constant of gravitation, and $P_{2k}(.)$ is the Legendre polynomial of degree $2k$. 

The spatial dependence of the ring potential is similar to the potential exterior to an axially symmetric spheroidal mass, such as that of a spheroidal sun in the axially symmetric approximation,  
\begin{equation}
  V_{\odot} = -\frac{G m_\odot}{r} \left[ 1 - \sum_{k=1}^\infty J_{2k}\Big(\frac{R_\odot}{r}\Big)^{2k} P_{2k} \left( \frac{z}{r} \right)\right],
\label{e:Vob}
\end{equation}
where $m_\odot$ is the solar mass, $R_\odot$ is the Sun's equatorial radius, and $J_{2k}$ are the coefficients of the zonal harmonics.
Then, provided that the ring plane is identified with the solar equator, comparing \eqref{e:Vob} with \eqref{e:Vring}, with both truncated to quadrupolar terms, we can define an ``effective $J_2$ of a hypothetical oblate Sun'' which approximately describes the orbit-averaged potential of a planet seen by a distant test particle,
\begin{equation}
  J_{2,\mathrm{eff}} = \frac{1}{2}\frac{m_\mathrm{p} a_\mathrm{p}^2}{m_\odot R_\odot^2} .
\label{e:J2eff}
\end{equation}
The values of $J_\mathrm{2, eff}$ arising from the inner three giant planets are given in Table~\ref{t:J2eff}.  The large values are owed to the very large ratio of orbit radius of the planets to the solar radius.
It is perhaps worth mentioning that the value of $J_{2,\mathrm{eff}}$ contributed by the terrestrial planets (Mercury, Venus, Earth and Mars) is only $\sim0.1$; this small value, as well as their total mass being less than $10^{-5}$ of the solar mass, justifies neglecting the terrestrial planets in the present analysis.

\begin{table}
\centering
\caption{Inner three giant planet parameters.}
\begin{tabular}{cccc}
\toprule
planet & $m_\odot/m_\mathrm{p}$ & $a_\mathrm{p}$ (au) & $J_\mathrm{2,eff}$ \\
\midrule
Jupiter  &    1047.3486  &    5.2076  &   592.5  \\
Saturn   &    3497.898   &    9.5725  &   605.5  \\
Uranus   &   22902.98    &   19.3038  &   376.1  \\
Total    &               &            &  1574.1  \\
\bottomrule
\end{tabular}

\addtabletext{%
Source for
 {$m_\odot/m_\mathrm{p}$}
and $a_\mathrm{p}$: \href{https://ssd.jpl.nasa.gov/horizons.cgi}{https://ssd.jpl.nasa.gov/horizons.cgi}, retrieved September 10, 2020.
}
\label{t:J2eff}
\end{table}

Below we report results of numerical simulations with a modified restricted three body problem of the Sun, Neptune and a massless Pluto {(\texttt{-{}-{}-NP})} in which we attributed an oblateness to the Sun. We represent the solar gravitational potential with a point-mass potential plus a second zonal harmonic with coefficient $J_2$
{(that is, up to $k=1$ in the series in \eqref{e:Vob})}.
We carried out a set of numerical simulations of this model for a time span of up to 2 Gyr, sampling 74 different values of $J_2$ in the range 1--10,000. 
 {The effect of the second zonal harmonic of the central mass} is implemented in the \textsc{swift} package that we use here.
Figure~{4} 
shows plots of the evolution of $\phi$ versus time and  {plots of $(e\cos\omega,e\sin\omega)$} 
for a selection of these simulations.
The results show that $\phi$ librates with a nearly steady amplitude of about $90^\circ$ in all cases, but the libration of $\omega$ with an amplitude below $45^\circ$ is found to persist only in a restricted range of $J_2$ of 1350--1650.
For $J_2\leq 500$ we find that $\omega$ circulates smoothly in a retrograde sense, and for $J_2\ge 3100$ it circulates smoothly in a prograde sense. In the boundary zones of $600\lesssim J_2 \lesssim 1300$ and $1700\lesssim J_2 \lesssim 3000$, $\omega$ has chaotic behavior, with intermittent librations and rotations; 
 {the inclination and eccentricity also have strongly chaotic behavior, correlated with each other and with that of $\omega$; two examples are shown in the Supplementary Information. In the zone of $600\lesssim J_2 \lesssim 1300$, the eccentricity can become large enough that the ensuing smaller perihelion distance would, in the actual solar system, allow closer approaches to Uranus and cause instability, such as found in the N-body simulations of the {\texttt{J-UNP}}, {\texttt{-SUNP}} and {\texttt{-{}-UNP}} models (cf.~Figure~\ref{f:nbodsB}).
}

\begin{figure*}
\centering
 \includegraphics[scale=0.9]{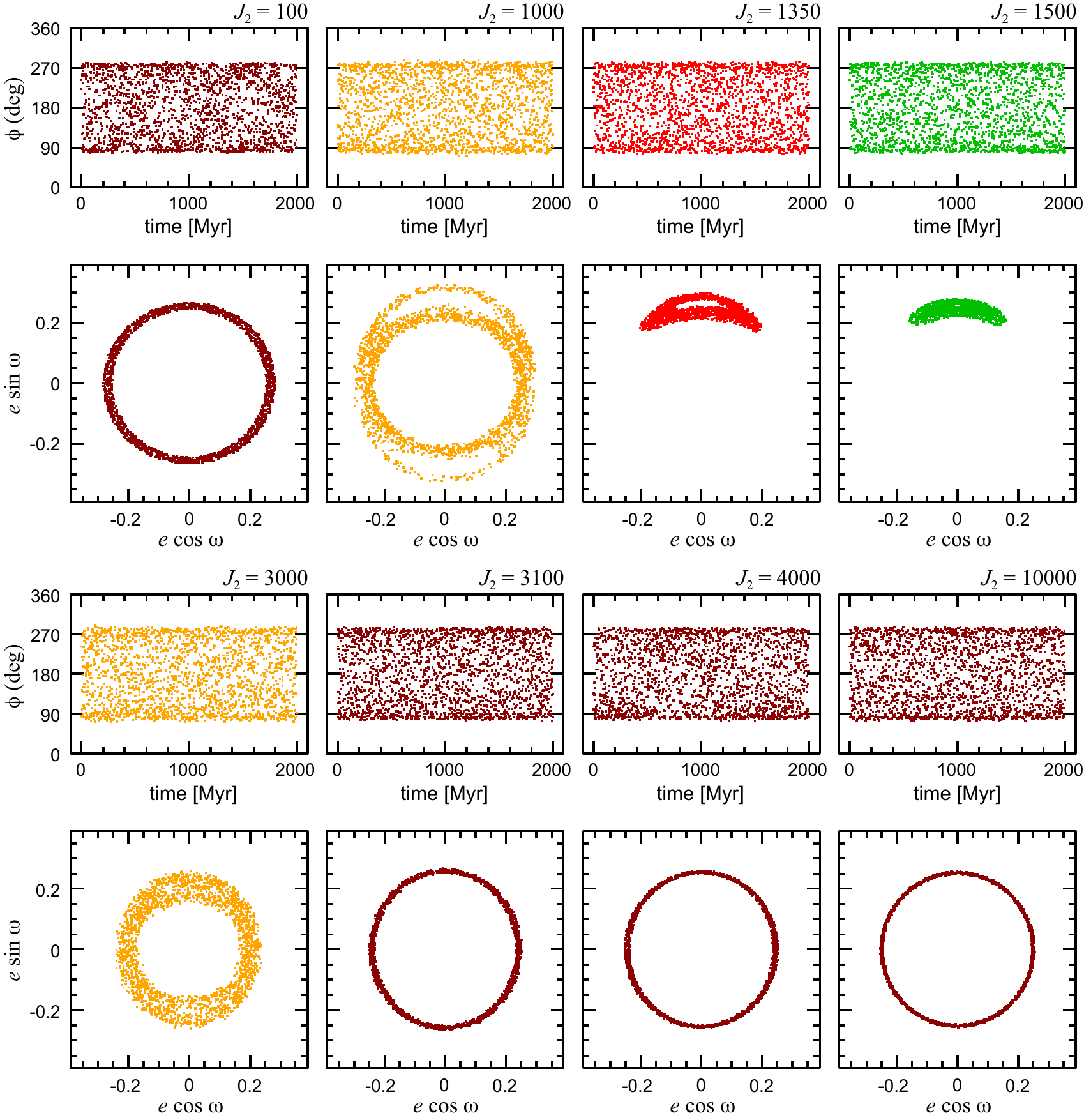}
\caption{
Numerical simulations for 2 Gyr of a modified three body model of the Sun, Neptune and (massless) Pluto 
in which we include  {an oblateness} 
for the Sun parameterized by a value of $J_2$, as indicated at the top of each pair of panels.
The top panels plot Pluto's critical resonant angle $\phi$ versus time, and the bottom panels are plots of $(e\cos\omega, e\sin\omega)$. Simulations were carried out for 74 values of $J_2$ in the range 1--10,000, only a small selection of 8 cases is shown here; see Supplementary Information for more detail.  The output interval in the plots is 1 Myr. The color scheme follows that of Figure~5. 
} 
\label{f:mod3bp}
\end{figure*}

\begin{figure*}[!t]\centering
 \includegraphics[scale=0.9]{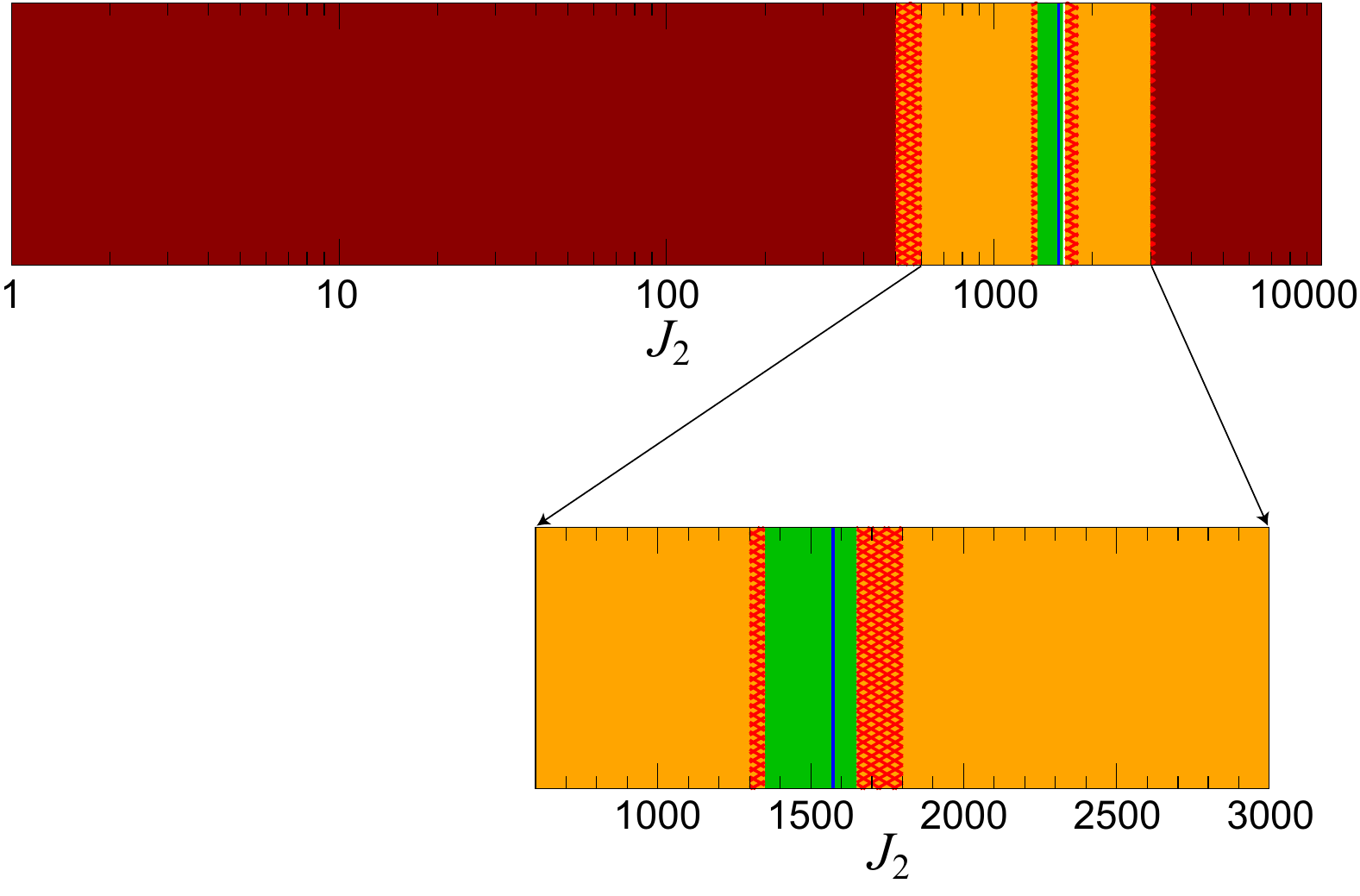}
\caption{
Summary of the behavior of Pluto's argument of perihelion {$\omega$} over 2 Gyr, as a function of the solar oblateness, $J_2$, in the modified restricted three body model.
In the dark red zone, $\omega$ rotates (in a retrograde sense in the small--$J_2$ regime, in a prograde sense in the large--$J_2$ regime).
In the orange zone it exhibits chaotic behavior with intermittent rotations and librations, and it librates steadily in the green zone.
The red hatched zones indicate the fuzzy boundaries between the neighboring zones. The blue vertical line indicates the value 1574.1 of the total effective $J_2$  estimated for the orbit-averaged combined quadrupolar effect of the inner three giant planets, Jupiter, Saturn and Neptune (see Table \ref{t:J2eff}).
Note that the horizontal scale in the upper panel is logarithmic from $J_2 =$ 1 to 10000, whereas in the lower panel it is linear from $J_2 =$ 600 to 3000.}
\label{f:fig5}
\end{figure*}

These results are summarized in Figure~5. 
 {The results support the hypothesis that the orbit-averaged perturbations of the three inner giant planets are the underlying physical mechanism that accounts for the latitudinal librations of Pluto's perihelion. They also highlight the narrow range of the effective $J_2$ for which librations of Pluto's $\omega$ are possible, and the remarkable circumstance that the orbital arrangement of the inner giant planets' yields an effective $J_2$ that happens to fall within this narrow range.}

Quantitatively, our result is somewhat different from that of Nacozy \& Diehl \cite{Nacozy:1978} who adopted the modified restricted three body model with the ``oblate Sun'' to carry out semi-analytic calculations; they reported an empirical estimate of $J_2=2005$ for the best agreement with Williams \& Benson \cite{Williams:1971}'s numerical solution for Pluto's motion on a timespan of $\sim4.5$~Myr. The difference is partly due to their semi-analytic approach versus our fully numerical approach, and partly due to updates in planetary masses and orbits that have occurred in the time since Nacozy \& Diehl's work. Notably, their estimate for the total effective $J_2$ lies in a range that our calculations find to be strongly chaotic for Pluto's argument of perihelion and its eccentricity and inclination (cf.~Figure~\ref{f:fig5}).

It is also interesting to note that in the modified restricted three body model, the libration of Pluto's critical resonant angle $\phi$ remains very stable for all the values of $J_2$ that we investigated. There is no indication of erratic evolution of $\phi$ of the kind found for two of the N-body  {models in Figure~\ref{f:nbodsA}, \texttt{-SUNP} and \texttt{-{}-UNP}}. For these two models, the equivalent modified three body problem would have total effective $J_{\mathrm{2}}$ values of 1198.0 and 376.1, respectively (see Table~\ref{t:J2eff}), and we would expect steady librations of $\phi$ based on the simulations of the modified three body model  {shown in Figure~\ref{f:fig5}}. From this comparison, we conclude that the origin of chaos in model \texttt{-SUNP} and model \texttt{-{}-UNP} lies not in the orbit-averaged secular perturbations of the three inner giant planets, but in their non-secular effects. A possible source of such non-secular effects of the inner giant planets is the near-resonant perturbations from Uranus:
Pluto's orbital period is close to three times as long as Uranus' orbital period, and
Neptune's orbital period is close to twice as long as Uranus' orbital period.
The timescale ($\lesssim1$~Myr) of the  {erratic} 
evolution found in models \texttt{-SUNP} and \texttt{-{}-UNP} in Figure~\ref{f:nbodsA}, is somewhat shorter than Pluto's secular apsidal and nodal precession timescales.
This also points to shorter timescale perturbations, as would arise from the near-resonant perturbations from Uranus.

\section*{Conclusions and future directions}

Tailored numerical simulations reported here elucidate the mechanisms underlying Pluto's perihelion librations and its long term dynamics. The results from these are summarized as follows.

\begin{enumerate}
    \item The stability of Pluto's perihelion librations in the azimuth (equivalently, the libration of the critical resonant angle, $\phi$) and in latitude (equivalently, the argument of perihelion, $\omega$) are sensitive to the perturbations of not only the most proximate planet (Neptune), but also to the inner giant planets (Jupiter, Saturn and Uranus).
    \item Neptune's influence is dominant in the libration of $\phi$.
    {However}, the other giant planets, particularly Uranus, influence the modulation of its amplitude of libration. 
    \item Uranus is the source of the most erratic perturbations. Without the stabilizing influence of Jupiter and Saturn, Uranus would destabilize the librations of both $\phi$ and $\omega$ on less than 10~Myr timescales. We conjecture that the reason for Uranus' destabilizing influence is its 3/1 near-resonance with Pluto, and possibly also indirectly its 2/1 near-resonance with Neptune.  
    \item 
     {It is rather striking that for Pluto-like orbits the architecture of the solar system's giant planets produces secular forcing of magnitude within the narrow range required to maintain the steady librations of $\omega$ (and corresponding steady variations of eccentricity and inclination) on gigayear timescales. This range is bounded by a zone in which $\omega$ undergoes strongly chaotic evolution (intermittent libration and rotation) and large chaotic variations of eccentricity and inclination on timescales much shorter than the age of the solar system}.
\end{enumerate}

Pluto's proximity to the edge of  {strong chaos in its latitudinal perihelion libration} invites further investigation. Better analytic approximations to assess the dynamical landscape in which Pluto orbits, including the effects of near-resonances with Uranus, are needed  { for further advancement of understanding of} both Pluto's dynamics and the dynamics of the large population of so-called Plutinos that orbit, alongside Pluto, in the same 3/2 resonance with Neptune \citep{Jewitt:1996,Yu:1999,Chiang:2002,Chiang:2003,Tiscareno:2009,Lawler:2013}. It is estimated that about 20\% of Plutinos share Pluto's property of a librating argument of perihelion \citep{Volk:2016}. 

In the previous literature, the latitudinal libration of Pluto's argument of perihelion is often called the von Zeipel--Lidov--Kozai (vZLK) oscillation, or the Lidov--Kozai oscillation or just the Kozai resonance \citep[e.g.][]{Milani:1989,Morbidelli:1995,Nesvorny:2000a,Gomes:2012}. Each of von Zeipel, Lidov and Kozai, independently predicted such a libration in some regimes of the orbit-averaged three body problem \citep{vonzeipel1910,lidov1961,Kozai:1962,Ito:2019}.  Additional perturbations such as the  quadrupole and higher order moments can significantly affect the vZLK oscillation \citep[e.g.][]{Tremaine:2014}. A quantitative and analytic formulation of the vZLK theory in the regime of mean motion resonances, such as Pluto's 3/2 mean motion resonance with Neptune, and including the secular effects of the inner giant planets, remains to be done; this topic is worth pursuing, and we intend to explore it in a forthcoming publication.

Improved understanding of Pluto's dynamics has broader implications for solar system dynamics. The orbital distribution of the Plutinos and other abundant resonant populations of minor planets beyond Neptune retain imprints of the dynamical history of the solar system. These imprints include the effects of resonance sweeping and capture by an outwardly migrating Neptune, and of the effects of gravitational scattering by the giant planets and possibly additional planets that may have existed briefly but were ejected from the solar system~\citep{Nesvorny:2018}. 
Prior to large-scale migration, the orbits of the giant planets would have been more compact and the effective $J_2$ would have been smaller. For magnitudes of migration considered in the recent literature, we calculate that the total effective $J_2$ (arising from Jupiter, Saturn and Uranus) would have remained within the range for Pluto's long term stability in its current orbit provided Uranus' outward migration was not more than about 5 au; this is 
discussed further in the Supplementary Information. We leave to future investigations to explore the implications of this limit for the migration history of the giant planets.
We can, however, state the general conclusion that, as a consequence of the large-scale migration of the giant planets, Pluto and the Plutinos were promoted into an orbital niche where minor planets can survive in eccentric and inclined orbits for multi-gigayear timescales, whereas their nearby dynamical neighborhood is strongly unstable. 
In our present state of understanding, Pluto's long term stability may be regarded as both inevitable and fortuitous, being owed in part to identifiable
physical mechanisms and in part to random processes inherent to those mechanisms.

Additional higher order resonances (including secular resonances between apsidal and nodal precession rates and so-called ``super-resonances'') have also been identified in Pluto's long term orbital dynamics~\citep{Milani:1989,Kinoshita:1996,Wan:2001}. These have periods exceeding $10^7$ years and cause only very small amplitude modulations of Pluto's perihelion \citep{Malhotra:1997}, and were not discussed in the present work. These weaker resonances may be relevant to explaining the weak chaos detected in numerical simulations of Pluto's long term motion \citep{Sussman:1992,Kinoshita:1996,Ito:2002}. Further investigations to examine the associated phase space regions and the role of these resonances would also help to probe the origin of the weak chaos.

In the present work,
Pluto's proximity to the edge of strong chaos has been determined within the specific model of the solar system, that is, with the current orbital architecture of the four giant planets as the only source of perturbations on Pluto's motion. Unmodelled perturbations could either increase or decrease Pluto's proximity to the edge of strong chaos. Therefore, Pluto's distance to the edge of strong chaos can potentially be used to quantify constraints on unmodelled perturbations that may have accumulated over its history, such as 
the collective gravity of the population of objects beyond Neptune as well as the effects of encounters or collisions with such objects,  {the effects of undiscovered distant planets,} and the perturbing effects of occasional close stellar flybys.  

Alternatively, we might question the implicit and widely held assumption that Pluto has remained 
in close proximity to its current orbit for much of the solar system's $\sim$~4.5 Gyr history, or at least since the end of the chaotic phase of formation and migration of the giant planets. Considering the results in this work on the proximity of Pluto to the edge of strong chaos, we must ask: could it be that Pluto's past orbital history on gigayear timescales is not as sanguine as assumed? We speculate that, with the inclusion of some types of unmodelled effects, it is perhaps not inconceivable that even in geologically recent times Pluto has an orbital history of intermittent chaotic episodes. 
The consequences of a chaotic orbital history would be significant for understanding Pluto's unexpected geophysical state, including the circumstances of its formation, the peculiar state of its spin axis and the properties of its satellite system \citep[e.g.,][]{Canup:2021}. Even if initially speculative, investigations along these lines may also identify geophysical evidence or dynamical arguments to either support or to rule out a chaotic orbital history.

\acknow{%
 {We thank the reviewers for helpful comments which improved the quality of this article.
Hiroshi Kinoshita gave us information about the history and current status of the analytic theory of Pluto's motion.
}
We also acknowledge NASA's Astrophysics Data System Bibliographic Services.
RM acknowledges research support from NSF (grant AST-1824869) and from the Marshall Foundation of Tucson, AZ. 
TI acknowledges research support from the JSPS Kakenhi Grant (JP18K03730/2018--2021).
The numerical orbit propagations were carried out at the Center for Computational Astrophysics (CfCA), National Astronomical Observatory of Japan.
}

\showacknow{%
} 

\section*{References}



\section*{Supplementary material}

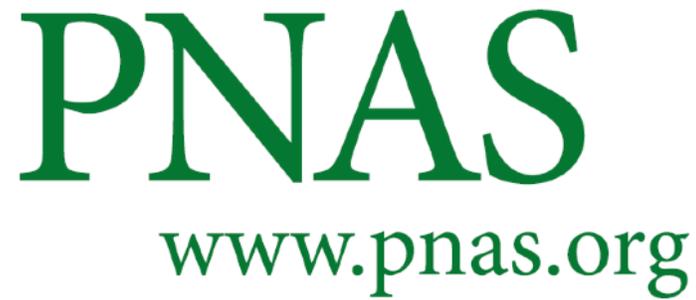

# Supplementary Information for

## Pluto near the edge of chaos

Renu Malhotra and Takashi Ito

Renu Malhotra
E-mail: malhotra@arizona.edu

**This PDF file includes:**

Supplementary text
Figs. S1 to S2 (not allowed for Brief Reports)
Tables S1 to S2 (not allowed for Brief Reports)
SI References

**Supporting Information Text**

**Eccentricity and inclination evolution**

In the simulations of the modified restricted three body model as well the full N-body model, the time variations of the eccentricity and the inclination of Pluto-like objects are both closely related to the time variations of the argument of perihelion, $\omega$. Figure S1 shows the time evolution of these three variables over 2 Gy for the highest fidelity N-body model, `JSUNP` (shown in blue, in the right-most panels), and for two cases of the modified restricted three body model (Sun–Neptune–Pluto, with $J_2 = 600$ and $J_2 = 1000$, shown in the left and middle panels). The latter two cases are illustrative of the range of $J_2$ values, 600–1300, in which a Pluto-like object's $\omega$ has chaotic behavior. In these cases we observe that the chaotic variations of the eccentricity and inclination have larger range than in the `JSUNP` system, and that the eccentricity reaches values exceeding $\sim 0.3$. Astronomical observations obtained thus far indicate that transneptunian objects in Neptune's 3/2 mean motion resonance are limited to eccentricities below $\sim 0.3$ (e.g. 1). The lower perihelion distance achieved with higher eccentricity leads to a larger destabilizing influence of closer approaches with Uranus. This explains the paucity of higher eccentricity Plutinos. This effect would be especially acute in the absence of steady latitudinal librations of the perihelion away from the plane of the planets. We therefore infer that in the regime of $600 \lesssim J_2 \lesssim 1300$, Pluto-like objects would be vulnerable to ejection from Neptune's mean motion resonance, with ensuing close approaches to the planets and eventual collision or transfer into scattered orbits or escape from the solar system.

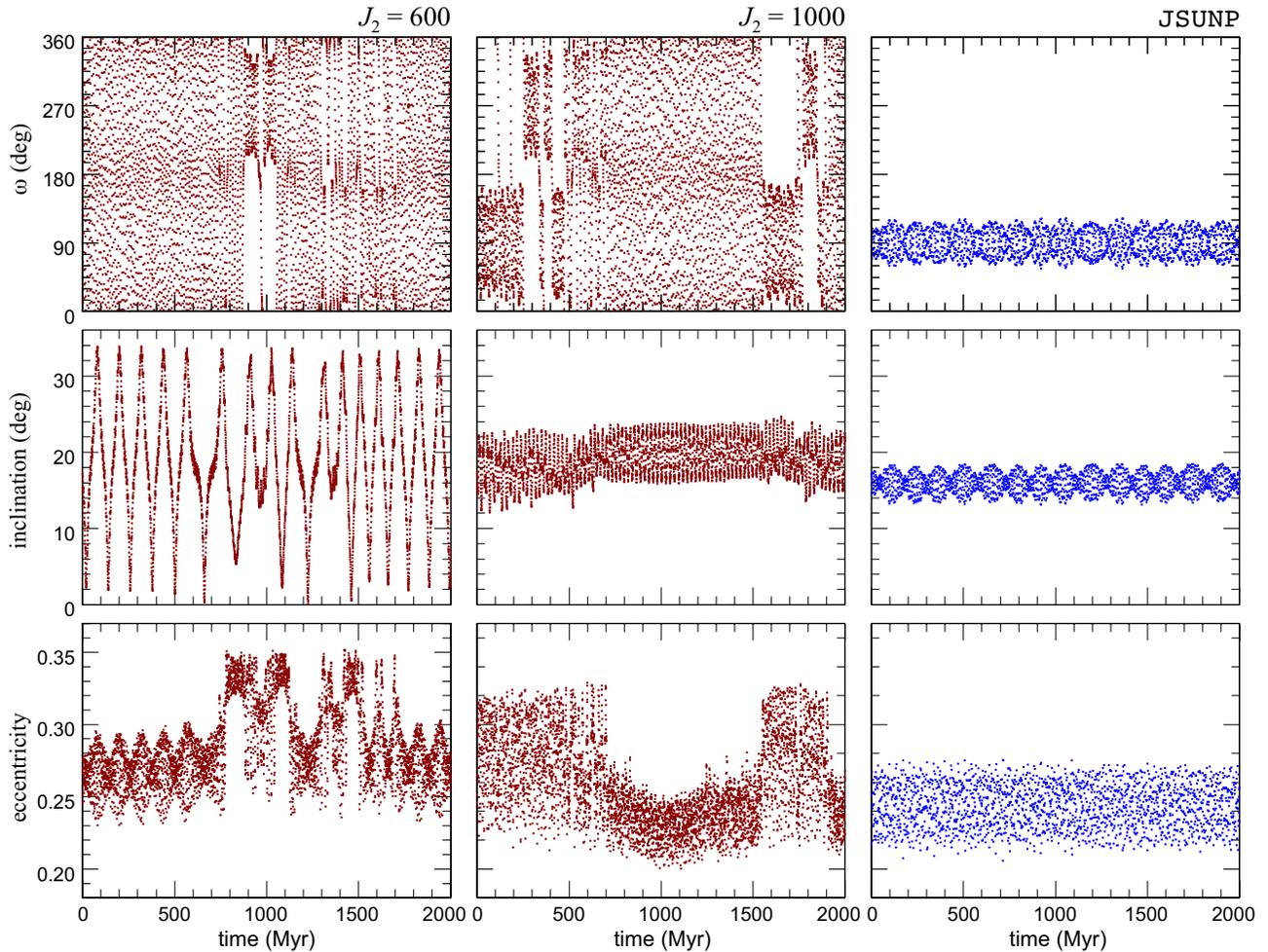

**Fig. S1.** Time evolution over 2 Gy of Pluto's eccentricity, inclination, and its argument of perihelion $\omega$, in three models: (left) modified restricted three body model (Sun–Neptune–Pluto) with total effective $J_2 = 600$, (middle) modified restricted three body model with total effective $J_2 = 1000$, (right) highest fidelity N-body simulation including all four giant planets (`JSUNP`).

**Effect of giant planets' migration on total effective $J_2$**

We can see from Eq. (4) in the main text that the time history of the total effective $J_2$ arising from Jupiter, Saturn and Uranus will depend on the time history of the semimajor axis of each of these giant planets during their past orbital migration. Currently, these time histories are not well constrained. (In principle, the total effective $J_2$ also depends upon the time history of the planetary masses, but it is reasonable to assume that the planetary masses did not change significantly after their formation.) In the previous literature, the magnitudes of migration discussed are approximately a few tenths au inward migration of Jupiter, $\sim 1$ au outward migration of Saturn, and a few au outward migration of Uranus. With these estimates, we find that the changes in the total effective $J_2$ during migration would have been most sensitive to Uranus' migration. As a first quantitative estimate, we find that the total effective $J_2$ would have remained within the range 1350–1650 that we found for Pluto's long-term stability in its current orbit (Figure 5 in the main text), provided that the magnitude of Uranus' outward migration was less than about 5 au. An outward migration of Uranus by more than $\sim 5$ au would potentially change the total effective $J_2$ enough to transition from the latitudinal libration of Pluto's argument of perihelion being unstable to stable. We note that better estimates would take account of the effects of the more massive ancient planetesimal disk and the changes in Pluto's orbit itself during the migration of the giant planets. Further investigation of this point may provide insights into the mechanism for capturing Pluto (and many Plutinos) into the libration state of the argument of perihelion that we observe today.

Figure S2 shows the time evolution of the total effective $J_2$ for a simple migration model in which the planets' orbits migrate with an $e$-folding timescale, $\tau$. The magnitude of each planet's migration is indicated in the figure legend.

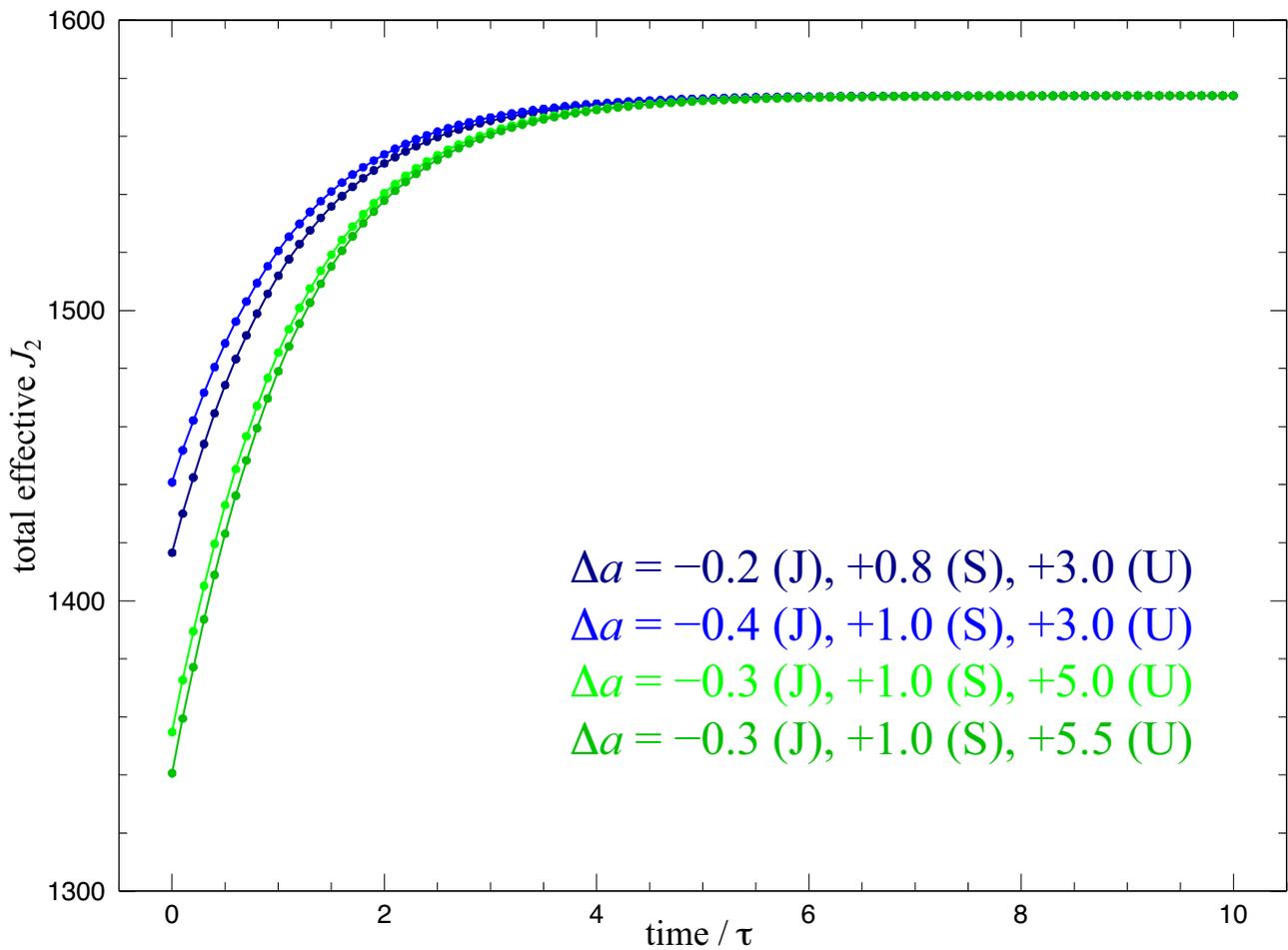

**Fig. S2.** Time evolution of the total effective $J_2$ in a simple model of the orbital migration of Jupiter, Saturn and Uranus. The planets are assumed to migrate with an $e$-folding migration timescale, $\tau$. The magnitude of the migration, $\Delta a$, of each planet is indicated in the legend. For example, "$-0.2$ (J)" means that Jupiter's semimajor axis decreases by $0.2$ au; "$+5.5$ (U)" means that Uranus' semimajor axis increases by $5.5$ au.

**Data Availability**

The data supporting this study's findings are available within the article and its Supplementary Information. Raw output data from our numerical orbit integrations is available from the authors upon reasonable request. See also Section Numerical experiments—N-body simulations in the main text for the details of the numerical integration scheme.

**Location of the SWIFT code package.** Access https://www.boulder.swri.edu/ hal/swift.html for retrieving the source code package.

**Masses of the Sun and the planets.** The masses of the Sun and the planets are listed in Table S1. The unit in this table is what is used in the SWIFT code where the gravitational constant is unity, the unit of length is au, and the unit of time is day, and one year is exactly 365.25 days. We adopted the masses of the planets from JPL's DE245 (e.g. 2). Note that Pluto is treated as a massless particle in our simulations, and it is not listed in Table S1.

**Table S1. Sun's and planet's masses.**

| object  | mass |
|---------|------|
| Sun     | $2.959139768995959 \times 10^{-4}$ |
| Jupiter | $2.82536279610815247 \times 10^{-7}$ |
| Saturn  | $8.45976574787474959 \times 10^{-8}$ |
| Uranus  | $1.29203489551819958 \times 10^{-8}$ |
| Neptune | $1.52436801162357287 \times 10^{-8}$ |

**Initial heliocentric positions and velocities of the planets and Pluto.** Table S2 shows the initial heliocentric positions and velocities of the planets and Pluto in our numerical simulations that we adopted from JPL Horizons System, and they are the values as of 2021 March 19 00:00:00 TDB (Barycentric Dynamical Time). The position and the velocity of each planet are those of its barycenter (including its satellite system). Pluto's position and velocity are also those of the barycenter of the Pluto system.

**Table S2. Initial positions $(x, y, z)$ and velocities $(v_x, v_y, v_z)$ of the planets and Pluto.**

| | | | |
|---|---|---|---|
| | | Jupiter | |
| $(x, y, z)$ au | $+3.479871576082048 \times 10^{0}$ | $-3.687143487014625 \times 10^{0}$ | $-6.254225671208241 \times 10^{-2}$ |
| $(v_x, v_y, v_z)$ au/day | $+5.402875942445465 \times 10^{-3}$ | $+5.540962998223180 \times 10^{-3}$ | $-1.438919301148455 \times 10^{-4}$ |
| | | Saturn | |
| $(x, y, z)$ au | $+5.820837611255400 \times 10^{0}$ | $-8.099314046939625 \times 10^{0}$ | $-9.084726091630656 \times 10^{-2}$ |
| $(v_x, v_y, v_z)$ au/day | $+4.225945590475764 \times 10^{-3}$ | $+3.247914368493194 \times 10^{-3}$ | $-2.246007629992486 \times 10^{-4}$ |
| | | Uranus | |
| $(x, y, z)$ au | $+1.515559431707809 \times 10^{1}$ | $+1.268157394646391 \times 10^{1}$ | $-1.492515199899297 \times 10^{-1}$ |
| $(v_x, v_y, v_z)$ au/day | $-2.547970714590798 \times 10^{-3}$ | $+2.838524448937736 \times 10^{-3}$ | $+4.340564757625335 \times 10^{-5}$ |
| | | Neptune | |
| $(x, y, z)$ au | $+2.950068586930330 \times 10^{1}$ | $-4.991750775881499 \times 10^{0}$ | $-5.771477267043819 \times 10^{-1}$ |
| $(v_x, v_y, v_z)$ au/day | $+5.087496282435223 \times 10^{-4}$ | $+3.120340880227267 \times 10^{-3}$ | $-7.590654143371072 \times 10^{-5}$ |
| | | Pluto | |
| $(x, y, z)$ au | $+1.428680876763813 \times 10^{1}$ | $-3.110962891479370 \times 10^{1}$ | $-8.023206165392421 \times 10^{-1}$ |
| $(v_x, v_y, v_z)$ au/day | $+2.933163077825810 \times 10^{-3}$ | $+6.417680020238108 \times 10^{-4}$ | $-9.148165866326484 \times 10^{-4}$ |

**Values of Sun's $J_2$.** The following values of Sun's $J_2$ are used for the simulations of the modified restricted three body problem whose results are summarized in Figure 5 in the main text: $J_2 = 1, 10, 100, 150, 200, 300, 400, 500, 600, 700, 800, 900, 1000, 1100, 1200, 1250, 1300, 1350, 1400, 1450, 1500, 1550, 1600, 1650, 1700, 1750, 1800, 1850, 1900, 1950, 2000, 2050, 2100, 2150, 2200, 2250, 2300, 2350, 2400, 2450, 2500, 2550, 2600, 2650, 2700, 2750, 2800, 2850, 2900, 2950 3000, 3100, 3200, 3300, 3400, 3500, 3600, 3700, 3800, 3900, 4000, 4100, 4200, 4300, 4400, 4500, 4600, 4700, 4800, 4900, 5000, 6000, 7000, and 10000. The time series of $\phi$ and the trajectories on the $(e\cos\varpi, e\sin\varpi)$ plane in eight of these cases, specifically $J_2 = 100, 1000, 1350, 1500, 3000, 3100, 4000, 10000$, are plotted in Figure 4 in the main text as illustrative examples.

**Specifying the $J_2$ values in the SWIFT code.** We assumed that the solar radius is 0.00465247265886874 au. This value becomes an input to calculate the effective $J_2$ values used in the SWIFT code. For example when $J_2 = 1000$, we enter $1000 \times 0.00465247265886874^2 = 0.02164550184152116$ in the pl.in parameter file for the SWIFT code.



\end{document}